\documentclass[preprint]{aastex61}

\usepackage{multirow}
\received{March 1, 2018}
\accepted{May 31, 2018}
\submitjournal{ApJ}

\shorttitle{Reproducing Type II WLF Observations with Electron and Proton Beams}
\shortauthors{Proch\'{a}zka et al.}

\begin{document}
\title{Reproducing Type II White-Light Solar Flare Observations with Electron and Proton Beam Simulations}
\correspondingauthor{Ond\v{r}ej Proch\'{a}zka}
\email{oprochazka01@qub.ac.uk}
\author{Ond\v{r}ej Proch\'{a}zka}
\affil{Astrophysics Research Centre, Queen's University Belfast, Northern Ireland, UK}
\author{Aaron Reid}
\affil{Astrophysics Research Centre, Queen's University Belfast, Northern Ireland, UK}
\author{Ryan O. Milligan}
\affil{Astrophysics Research Centre, Queen's University Belfast, Northern Ireland, UK}
\affil{SUPA School of Physics and Astronomy, University of Glasgow, Scotland, UK}
\affil{NASA Goddard Space Flight Centre, Greenbelt, MD 20771, USA}
\affil{Department of Physics, Catholic University of America, 620 Michigan Avenue, Northeast, Washington, DC 20064, USA}
\author{Paulo J. A. Sim\~{o}es}
\affil{SUPA School of Physics and Astronomy, University of Glasgow, Scotland, UK}
\author{Joel C. Allred}
\affil{NASA Goddard Space Flight Centre, Greenbelt, MD 20771, USA}
\author{Mihalis Mathioudakis}
\affil{Astrophysics Research Centre, Queen's University Belfast, Northern Ireland, UK}

\begin{abstract}

We investigate the cause of the suppressed Balmer series and the origin of the white-light continuum emission in the X1.0 class solar flare on 2014 June 11. We use radiative hydrodynamic simulations to model the response of the flaring atmosphere to both electron and proton beams which are energetically constrained using RHESSI and Fermi observations. A comparison of synthetic spectra with the observations allow us to narrow the range of beam fluxes and low energy cut-off that may be applicable to this event. We conclude that the electron and proton beams that can reproduce the observed spectral features are those that have relatively low fluxes and high values for the low energy cut-off. While electron beams shift the upper chromosphere and transition region to greater geometrical heights, proton beams with a similar flux leave these areas of the atmosphere relatively undisturbed. It is easier for proton beams to penetrate to the deeper layers and not deposit their energy in the upper chromosphere where the Balmer lines are formed. 
The relatively weak particle beams that are applicable to this flare do not cause a significant shift of the $\tau=1$ surface and the observed excess WL emission is optically thin.  

\end{abstract}


\section{Introduction} 
Despite a large number of white-light flare (WLF) observations, the processes that deliver energy to the deepest layers of the Sun's atmosphere where optical emission is believed to be formed, are poorly understood (\citealt{Hudson:2016aa}). \cite{Watanabe:2010aa} and \cite{Watanabe:2017aa} reported observations of WLFs, but were unable to conclude how the required energy was delivered to the deeper layers of the atmosphere. \cite{Fletcher:2007ab} analyzed WLF observations from the Transition Region and Coronal Explorer (TRACE; \citealt{Handy:1999aa}) and the Ramaty High Energy Solar Spectroscopic Imager (RHESSI; \citealt{Lin:2002aa}) of WLFs with classifications ranging from C4.8 to M9.1. They estimated that an electron beam with a low energy cut-off below 25~keV could carry sufficient energy, but concluded that electrons with such low energies can not penetrate sufficiently deep into the lower solar atmosphere. X-ray spectroscopy of an X1 class flare showed an unusually high low energy cut-off of $\approx$ 100~keV which may explain the lack of substantial energy deposition in the chromosphere (\citealt{Warmuth:2009aa}). Alfv\'{e}n waves are another possible mechanism that can explain the rapid energy transport from the corona to the lower solar atmosphere during the impulsive phase of flares (\citealt{Fletcher:2008aa,Kerr:2016aa,Hao:2017aa,Reep:2018aa}).

The wide range of WLFs that have been observed to date can be grouped into two main categories. Type I WLFs show the presence of Balmer and Paschen edges and have more intense emission, while the events with significantly weaker hydrogen emission lines and a relatively flat continuum are grouped into Type II (\citealt{Machado:1986aa}, \citealt{Canfield:1986aa}). Observations of Type II WLF were first presented by \cite{Boyer:1985aa}. Their spectral analysis, which assumed optically thin emission, ruled out an origin of the WL emission as a result of the Paschen continuum. Optically thin H$^-$ emission would imply intense heating of the lower atmosphere but requires a process that would deposit significant amounts of energy in the deeper layers. Their calculations have shown that this deposition of energy would be accompanied with an increase in temperature by $\approx$2000~K at a height of $\approx$200~km above the photospheric floor or an increase in temperature by $\approx$150~K over the entire photosphere and chromosphere. Based on observations of three WLFs, 
\cite{Fang:1995aa} concluded that besides the observed spectral differences, a temporal mismatch between the WL and HXR emission is also an indication of Type II WLFs. \cite{Potts:2010aa} associated the temporal and spatial correspondence of WL emission and hard X-ray sources with the so-called thick-target model. Assuming that the WL excess emission originated above the photosphere, their analysis of an X3-class flare concluded that the WL excess emission was optically thin. 

Some of the scenarios used to explain Type II WLF include photospheric reconnection, radiative back-warming, and particle beams that are able to penetrate through the chromosphere. Photospheric reconnection would be most efficient at the temperature minimum region and result in localized heating in the photosphere (\citealt{Li:1997aa}, \citealt{Chen:2001aa} and \citealt{Litvinenko:1999aa}). \cite{Ding:1999aa} modeled the flare continuum emission using a high-energy particle beam accelerated in the temperature minimum region. This led to an initial decline in intensity (so-called \textit{black-light flare}), absence of the Balmer discontinuity and only a minor disturbance in the chromosphere where the Balmer lines are formed. \cite{Machado+89} estimated that electrons would require an energy of at least 170~keV to reach the temperature minimum region (TMR), meanwhile protons would need energies of 6~MeV. They concluded that the electron energy is too high and given the lack of observational evidence for protons the particles must be stopped in the chromosphere where the Balmer continuum is formed. Balmer continuum radiation (through the process of back-warming) could provide the energy for heating the photosphere leading to H$^-$ emission and a WL continuum. \cite{Allred:2005ab} used more sophisticated chromospheric heating models and found that back-warming contributes only 10\% to the heating. In a recent study \cite{Simoes:2017aa} used RADYN simulations to determine the formation of the infrared continuum in flares and found no enhancement in the photospheric blackbody emission.

The scenario of flare energy transfer by proton beams was first proposed by \cite{Svestka:1970aa}, who calculated a threshold of 20~MeV as the lowest energy required by protons to penetrate into the upper photosphere. \cite{Simnett:1986aa} highlighted that the traditional flaring scenario which employs electron beams with a low energy cut-off of 20--25~keV is not consistent with many observations and favored shock accelerated protons to explain the thermal X-ray bursts at the beginning of the impulsive phase.  More recent observations by \cite{Martinez-Oliveros:2012aa} found both WL and HXR sources in photosphere during the impulsive phase of a flare. However, these observations cannot be explained with any plausible electron beam scenario as 20--25~keV electrons cannot penetrate to heights of $\approx$195~km above the photospheric floor. 
In recent years RHESSI has been used to obtain electron beam parameters which are then used as an input into to radiative hydrodynamic simulations. The compression of the lower atmosphere, as a result of the electron beam heating, allows for higher energy electrons to penetrate to lower altitudes. The location of the  deepest penetration coincides with the peak in the contribution function of the Balmer continuum (\citealt{Kennedy:2015aa}). 

\cite{Prochazka:2017aa} reported the observation of an X1 WLF that was observed at the Ond\v{r}ejov Observatory, Czech Republic, with the \textit{Image Selector} (IS, \citealt{Kotrc:2016aa}) instrument, providing rare optical spectra (spectral resolution $\sim$0.03 nm per pixel) in conjunction with modern space-based instruments. The authors reported no emission in the higher order Balmer lines, as well as weak 
emission in Lyman lines and Lyman continuum (LyC). They compared these observations with synthetic line profiles generated by two distinct heating models: one using a generic electron beam, and one where the heating was deposited directly in the TMR. The deposition of energy in the TMR led to an increased optical continuum 
and only weak emission in the wings of the Balmer lines and the Balmer jump, in broad agreement with the observations. The continuum generated by the model included contributions from both blackbody emission and Balmer continuum. 
Their analysis concluded that depositing the energy deep in the atmosphere can lead to increased continuum and a suppression of the hydrogen line emission. Although electron beams can not be excluded from the interpretation of the observations, the parameters of the beam must be rather extreme.

In this paper we carry out a more detailed analysis of electron and proton beam heating models in order to explain the observations of the type II WLF presented by \cite{Prochazka:2017aa}. In Section~\ref{flare_obs} we provide an outline of the datasets obtained from both ground- and space-based instruments. In Section~\ref{radyn_section} we model the response of the lower solar atmosphere to both electron and proton beams using 1D radiative hydrodynamics. In Sections~\ref{results_section} we present our findings, with a discussion presented in Section~\ref{disc_section}. The conclusions are summarized in Section~\ref{conc_section}.

\section{The X1 flare on 2014 June 11: observations and data analysis}
\label{flare_obs}

\begin{figure}[!t]
\centering
\includegraphics[width=\textwidth]{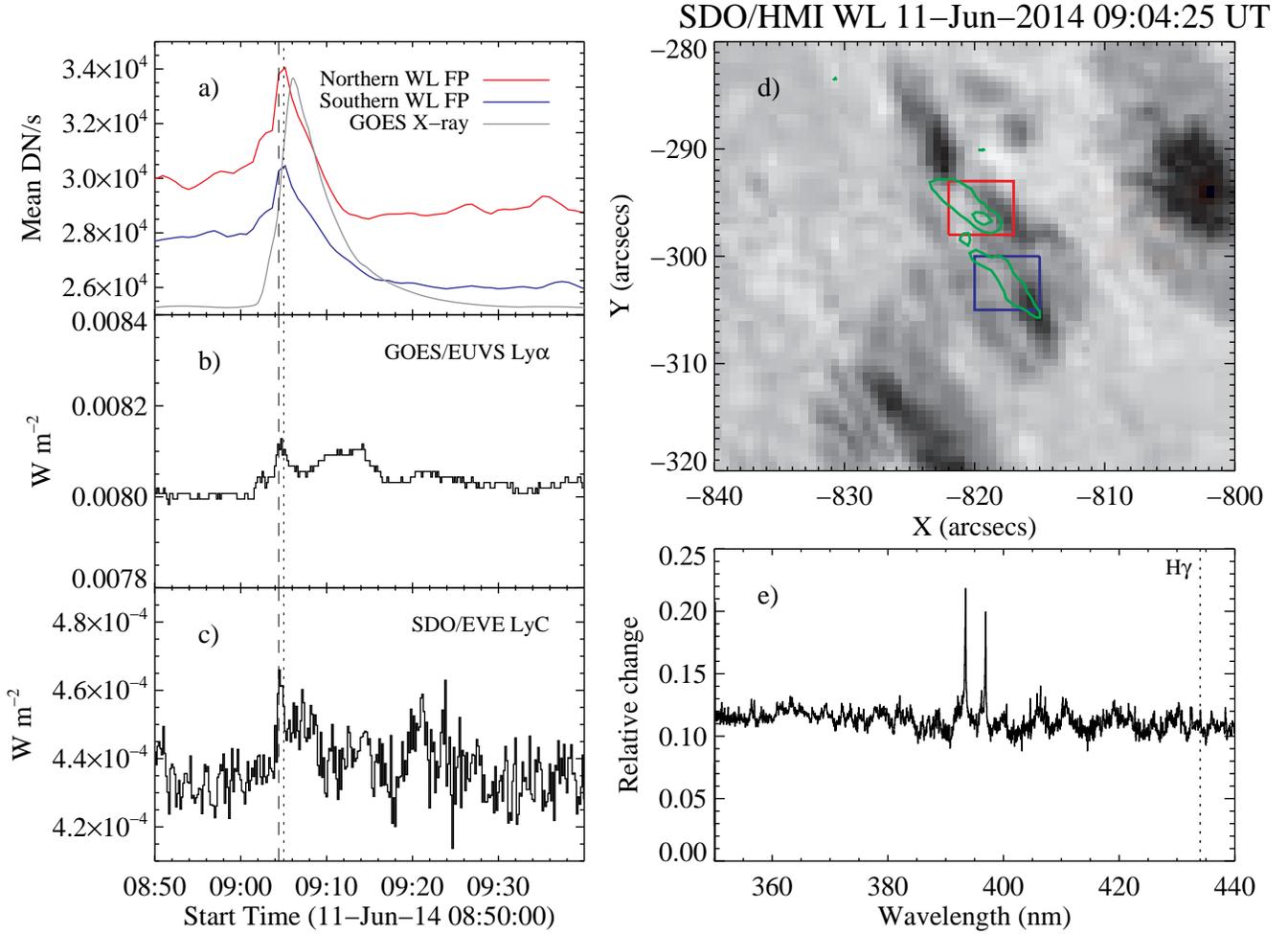}
\caption{Summary of observations taken during the 2014 June 11 flare. a) lightcurves of WL emission from the northern (red) and southern (blue) footpoints as highlighted in panel d). GOES 1--8\AA\ time profile is also shown for reference. b) Ly$\alpha$ lightcurve from GOES/EUVS. c) LyC lightcurve from SDO/EVE. The vertical dashed and dotted lines in panels a--c denote the times of the WL image (in panel d) and IS spectrum (in panel e), respectively. d) SDO/HMI WL image showing the location of two WL kernels. e) IS flare excess spectrum relative to a pre-flare profile (averaged over 09:04:45--09:05:15~UT). The reference spectrum was recorded at 08:54:45--08:55:15~UT.}
\label{wl_flare_fig}
\end{figure}


On 2014 June 11, an X1.0 WLF (that peaked at 09:06~UT) in active region NOAA 12087 (location S18E57) was observed by the IS instrument, which provides high-temporal resolution spectroscopy (10 spectra per second) in the $\lambda$~=~350--485~nm wavelength range, as well as H$\alpha$ context images. The IS spectra are integrated over an area of $\sim$1 arcmin in diameter. Using the H$\alpha$ images we estimate the flaring kernels to cover 7\% 
of this area, indicating a small filling factor of $f\!f=0.07$. Due to the nature of the observations the noise may mask any weak Balmer line emission. We quantify these effects in the spectrum of Figure 1 by measuring the height of the Ca II K line (from the continuum) with respect to the flux at the expected location of H$\gamma$. Based on this analysis we conclude that the H$\gamma$ vs. \ion{Ca}{2}~K line ratio must be lower than 0.1.
Errors in our measurements reached 0.5\% for wavelengths $\sim$440~nm and integration times of 30~ms and up to 2\% for the \ion{Ca}{2}~K\&H lines and wavelengths of $\sim$360~nm. The errors were determined as the standard deviation over a set of 50 reference spectra. The observations in visible range showed that the higher Balmer lines remained in absorption while the H$\alpha$ intensity showed a clear increase during the flare (Figure~1 in \citealt{Prochazka:2017aa}). The higher order Balmer lines (e.g. H$\gamma$) do not show the characteristically strong emission that may be expected during the impulsive phase.  Any systematic rise both redward and blueward of the Balmer jump was not detected during the impulsive phase (Figure~\ref{wl_flare_fig}e). It should be noted that the IS spectrum shown in Figure~\ref{wl_flare_fig}e is different from those presented in \cite{Prochazka:2017aa}, as, in this work, the reference pre-flare spectrum was taken much closer to the flare onset (08:54:45--08:55:15~UT).
Observations by the EUV Sensor on \textit{Geostationary Operational Environmental Satellite} (GOES/EUVS; \citealt{Viereck:2007aa}) and the \textit{SDO/EUV Variability experiment} (SDO/EVE; \citealt{Woods:2012aa}) showed a similar behavior in the Lyman series, with weak emission  in the Ly$\alpha$ line (Figure~\ref{wl_flare_fig}b) and LyC (Figure~\ref{wl_flare_fig}c), respectively.

\subsection{White-light from SDO/HMI Observations}
\label{hmi_section}

The WL continuum emission near the \ion{Fe}{1} 617.3~nm line recorded by the \textit{Helioseismic and Magnetic Imager} (SDO/HMI, \citealt{Scherrer:2012aa}) peaked at the same time and was co-spatial with the hard X-ray source (Figure~\ref{wl_flare_fig}a and d; see also Figure~2 in \citealt{Prochazka:2017aa}). We were only able to reliably determine a lower limit of the WL contrast in the observations due the large temporal and spatial variations in the active region (Figure~\ref{wl_flare_fig}a). For the northern and southern kernels the WL contrast was $\geq$1.07 and $\geq$1.05
 respectively. We used HMI continuum images to estimate the area of the WL emission. We applied intensity thresholding on difference images. This allowed us to select upper and lower limits of the flaring area. 
If we consider the location of the flare (S18E57), the geometric distortion would cause the observed flaring area to look about 2.5 times smaller. The resulting area was in a range 1.1$\times$10$^{17}$--3.3$\times$10$^{17}$ cm$^2$. This area, combined with the power of the non-thermal electrons (Section \ref{rhessi_section}), provided lower and upper estimates on the energy flux for both the electron and proton beams. 

\subsection{Hard X-rays from RHESSI Observations}
\label{rhessi_section}

\begin{figure}[p]
\centering
\includegraphics[width=0.84\textwidth]{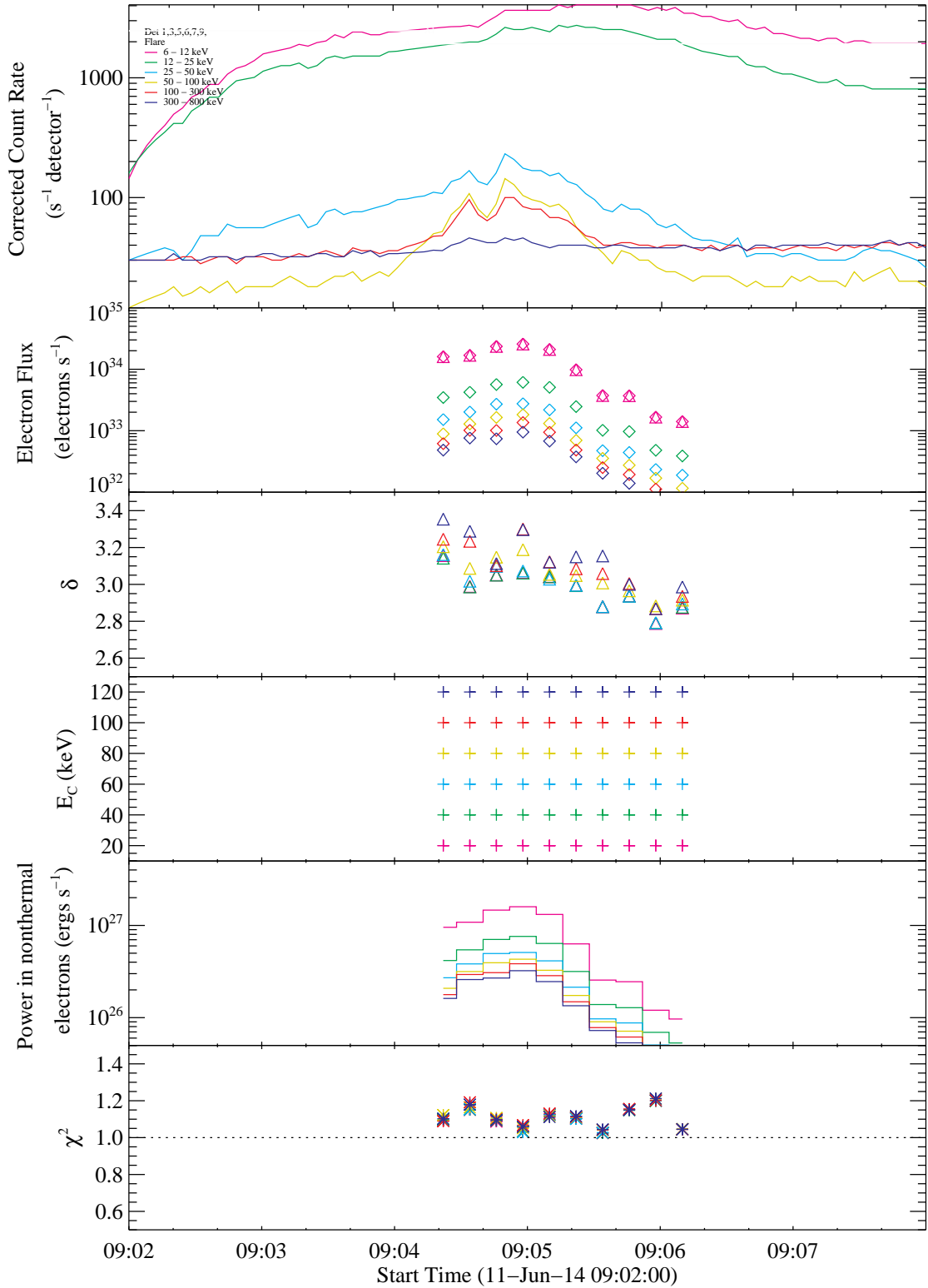}
\caption{RHESSI fitting results with low energy cut-off fixed (color coded) during the impulsive phase. First panel: Corrected Count Rate, second panel: Electron flux, third panel: Spectral index, forth panel: Low energy cut-off, fifth panel: Power in non-thermal electrons, sixth panel: Chi-squared.}
\label{hsi_params_fig}
\end{figure}

RHESSI observed the flare in the time interval 8:18 UT to 9:21~UT. The RHESSI light curve shows that the higher energy bands peaked at 9:04:45~UT. X-ray spectra were generated for ten 12~s intervals across the flare peak using only detector \#7 (Figure~\ref{hsi_params_fig}). At this stage of the mission, the detectors had suffered severe degradation and detector \#7 was found to have the highest sensitivity out of the 9 collimators. Note that the RHESSI detectors were annealled for the fourth time between 2014 June 26 and 2014 August 13; just 15 days after this flare occurred.

Initial attempts to fit the RHESSI spectra with a combination of a multi-thermal component at lower energies and a non-thermal component at higher energies, (as well as the standard albedo, pulse-pile up and detector response matrix corrections), failed to provide consistent estimates for the low energy cut-off ($E_C$). Ordinarily, hard non-thermal spectra are often easier to fit than softer spectra as the non-thermal tail deviates more significantly from the thermal Maxwellian distribution. However, as this flare exhibited an unusually hard slope ($\delta$ $\sim$ 3), the flattening of the photon spectrum below $E_C$ was similar to the slope above $E_C$ making it difficult to distinguish between different values of $E_C$. To this end the fitting process was performed with fixed values of $E_C$ at 20, 40, 60, 80, 100 and 120~keV, while the slope and normalization factor (i.e. the number of electrons) were allowed to vary. The quality of the fit, represented with $\chi^2$, turned out to be independent of the value of E$_C$ (Figure~\ref{hsi_params_fig}).

The value of $E_C$ is crucial for calculating the total power of non-thermal electrons, because we assume that the electron distribution above $E_C$ is given by a power-law function, meanwhile it is equal to zero below this value (\citealt{Holman:2011aa}). For a fixed amount of energy in non-thermal electrons, a low value of $E_C$ leads to a high power in non-thermal electrons above $E_C$ and vice versa. Having obtained the power in non-thermal electrons from  the fits to the RHESSI spectra for each $E_C$, the flux (in erg cm$^{-2}$ s$^{-1}$) was found by dividing by the WL area derived from HMI data, as described in Section~\ref{hmi_section}.

\begin{deluxetable}{ccc}
\tablecaption{Estimates of the power of non-thermal electrons in the impulsive phase of the flare and the derived maximum flux for given values of the low energy cut-off with respect to a flaring area in the range of 1.1$\times$10$^{17}$ to 3.3$\times$10$^{17}$~cm$^{2}$. \label{tabX}}
\startdata
\\
E$_C$ (keV)  &  Power (erg s$^{-1}$) & Maximum flux (erg cm$^{-2}$ s$^{-1}$) \\
\hline
20 & 1.6$\times$10$^{27}$ & 4.85 - 14.5$\times$10$^{9}$  \\
40 & 7.6$\times$10$^{26}$  & 2.30 - 6.91$\times$10$^{9}$ \\
60 & 5.1$\times$10$^{26}$  & 1.55 - 4.64$\times$10$^{9}$  \\
80 & 4.3$\times$10$^{26}$  & 1.30 - 3.90$\times$10$^{9}$  \\
100 & 3.8$\times$10$^{26}$  & 1.15 - 3.45$\times$10$^{9}$  \\
120 & 3.2$\times$10$^{26}$ & 0.97 - 2.91$\times$10$^{9}$  \\
\enddata
\end{deluxetable}

\subsection{$\gamma$-rays from Fermi/GBM Observations}
\label{fermi_section}

\begin{figure}[h]
\centering
\includegraphics[width=0.6\textwidth,angle=270]{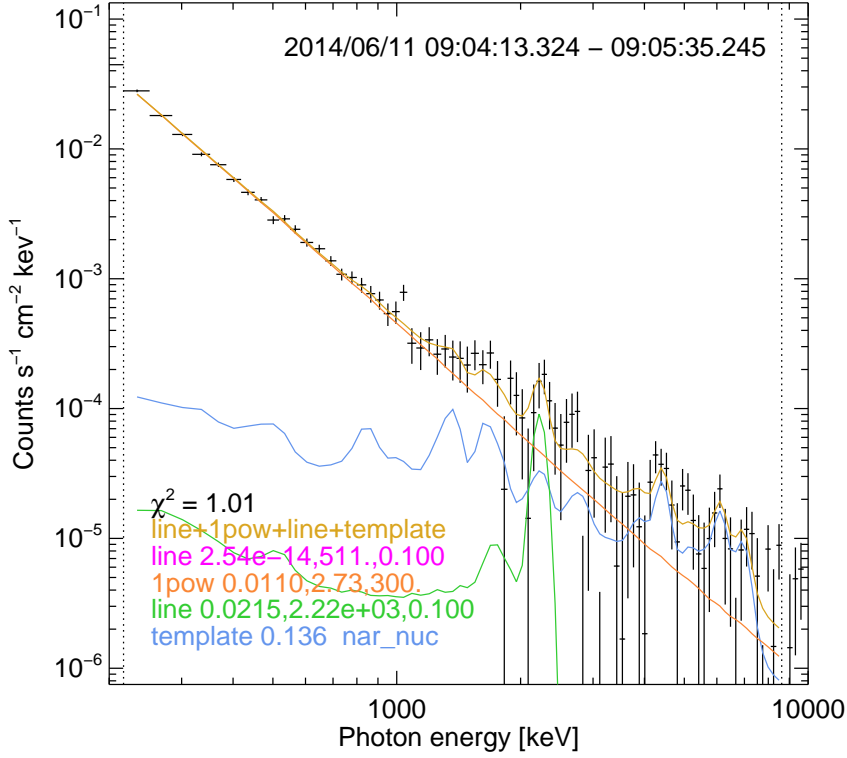}
\caption{Fermi/GBM count spectrum in the range 200~keV--10~MeV from the BGO detector, integrated between 09:04:13 and 09:05:35~UT. The spectrum was fitted with a power-law (orange), 511~keV line (magenta), 2.223~MeV line (green), and nuclear line template (blue), see text for details. The vertical dotted-lines indicate the energy range used for the spectral fitting.}
\label{fig:bgo}
\end{figure}

The {\it Gamma Ray Burst Monitor} (GBM) on board the \textit{Fermi Gamma-ray Space Telescope} (\citealt{Meegan:2009aa}) detected sufficient counts in the 200--10,000~keV range to allow an estimate of the parameters of the accelerated protons. The background counts were determined by using a linear fit to the counts before (between 08:56:08 and 09:01:20~UT), and after (09:08:39 and 09:21:47~UT) the flare. After subtracting the background, the bismuth germanite detector (BGO) counts were integrated between 09:04:13 and 09:05:35~UT to produce the spectrum. The spectrum was fitted with a power-law to capture the electron bremsstrahlung component, Gaussians for the 511~keV electron-positron annihilation line and 2.223~MeV neutron capture lines, and a template to describe the narrow de-excitation nuclear lines (Figure~\ref{fig:bgo}). The template for nuclear lines is included as standard in OSPEX \citep{Schwartz:2002aa}, and it is calculated for a flare at a heliocentric angle of 60$^{\circ}$ by assuming a downward isotropic distribution of ions and a power-law energy distribution with spectral index 4 and an $\alpha$/p ratio of 0.22. The template is normalized so that a value of 1 photon s$^{-1}$ cm$^{-2}$ keV$^{-1}$ corresponds to $8.5946 \times10^{29}$ protons per second with energies above 30~MeV (\citealt{Trottet:2015aa}). Thus, the total number of accelerated ions above 30 MeV ($N_{E_p>30MeV}$) is proportional to the template normalization.
We remark that the template was generated by an ion distribution with $E_C=1$~MeV. In order to obtain the total number of ions ($N_{E_p>E_C}$) above a given cutoff $E_C$ it is then necessary to extend the power-law ion distribution (with index 4) down to a lower energy cutoff (here, 2 MeV) to account for the protons in the energy range required to trigger the nuclear reactions (2 to 10 MeV), as defined by the cross-section for such interactions (\citealt{Murphy:2007aa,Vilmer:2011aa}).

The $\gamma$-ray emission was rather weak above 2.5~MeV; we have included the counts above this energy only to estimate an upper limit for the fitting. The results of the spectral analysis are presented in Table \ref{tabFermi}. These results average the proton numbers and their energy for the entire duration of the event, giving a lower limit for $N_{E_p}$. However, it is clear that these values are not constant for the entire period. Therefore, we repeated the procedure for a shorter time interval, 09:04:29 to 09:04:50~UT (but only fitting the spectrum below 2~MeV due to poor count statistics) in order to estimate the proton number and energy closer to the peak of the impulsive phase of the flare. Note that the total number of ions $N_{30}=N_{E_p>30MeV}\Delta t$ obtained from both integration times are consistent within their uncertainties, indicating that the majority of the accelerated ions were accelerated within the 21~s window at the peak of the impulsive phase. The results are also shown in Table~\ref{tabFermi}. We estimate the maximum flux in proton beams for the values of $E_C$= 2, 4, 8 and 16~MeV (Table \ref{tabXF}) following a procedure similar to the one applied to the RHESSI spectra.

\begin{deluxetable}{ccc}
\tablecaption{Fermi/GBM BGO spectral results for the impulsive phase. \label{tabFermi}}
\startdata
\tablehead{Integration interval &  09:04:13 to 09:05:35~UT  & 09:04:29 to 09:04:50~UT \\ Integration time $\Delta t$ & 82~s & 21~s}
line 511~keV & $\sim 0$ (no detection) & 0.023 $\pm$ 0.026 ph~s$^{-1}$~cm$^{-2}$ (upper limit)\\
line 2.223~MeV & 0.02 $\pm$ 0.01 ph~s$^{-1}$~cm$^{-2}$ & 0.004 $\pm$ 0.021 ph~s$^{-1}$~cm$^{-2}$\\
power-law normalization & 0.011 ph~s$^{-1}$~cm$^{-2}$~keV$^{-1}$ at 300~keV & 0.017 ph~s$^{-1}$~cm$^{-2}$~keV$^{-1}$ at 300~keV\\
photon spectral index & 2.73 $\pm$ 0.02 & 2.73 $\pm$ 0.02\\
Nuclear lines template  & 0.14 $\pm$ 0.02 ph~s$^{-1}$~cm$^{-2}$ & 0.51 $\pm$ 0.09 ph~s$^{-1}$~cm$^{-2}$\\
$N_{E_p>30\mathrm{MeV}}$ & $1.2\pm 0.2 \times10^{29}$~ions~s$^{-1}$ & $4.4 \pm 0.8 \times10^{29}$~ions~s$^{-1}$\\
{\bf $N_{30}$} & $9.8 \pm 1.6 \times 10^{30}$ ions & $9.2 \pm 1.7 \times 10^{30}$ ions \\ \hline
\multicolumn{3}{c}{Parameters calculated from the fitting results}\\
$P_{E_p>30\mathrm{MeV}}$ & $8.3 \times10^{24}$~erg~s$^{-1}$ & $3.1 \times10^{25}$~erg~s$^{-1}$\\
$N_{E_p>2\mathrm{MeV}}$ & $3.9 \times10^{32}$~ions~s$^{-1}$ & $1.5 \times10^{33}$~ions~s$^{-1}$\\
$P_{E_p>2\mathrm{MeV}}$ & $1.9 \times10^{27}$~erg~s$^{-1}$ & $7.1 \times10^{27}$~erg~s$^{-1}$\\
\enddata
\end{deluxetable}

\begin{deluxetable}{ccc}
\tablecaption{Estimates of power in non-thermal protons during the impulsive phase of the flare and the derived maximum flux for given values of the low energy cut-off with respect to a flaring area in the range of 1.1$\times$10$^{17}$ to 3.3$\times$10$^{17}$~cm$^2$. \label{tabXF}}
\startdata
\tablehead{E$_C$ (MeV) &  Power (erg s$^{-1}$)  & Maximum flux (erg cm$^{-2}$ s$^{-1}$)}
2 & 7.06$\times$10$^{27}$  & 21.4--64.2$\times$10$^{9}$  \\
4 & 8.82$\times$10$^{26}$  & 2.67--8.02$\times$10$^{9}$  \\
8 & 1.10$\times$10$^{26}$  & 0.33--1.00$\times$10$^{9}$  \\
16 & 1.38$\times$10$^{25}$  & 0.04--0.13$\times$10$^{9}$  \\
\enddata
\end{deluxetable}

\section{RADYN modeling}
\label{radyn_section}

The RADYN code \citep{Carlsson:1992aa,Carlsson:1995aa,Carlsson:1997aa,Allred:2015aa} was used to model the response of the solar atmosphere to a set of plausible heating parameters. RADYN solves the equations of radiative hydrodynamics in one dimension with an adaptive grid and allows for direct thermal heating and/or a particle beam to be applied. The initial model used in this work is for a plage-like atmosphere (QS.SL.HT from \citealt{Allred:2015aa}). It assumes a 10~Mm half loop with a reflected top boundary and a coronal temperature of 3~MK (\citealt{vernazza:1981aa}). The transition region is placed $\sim$1300~km above the photospheric floor, to mimic the more active atmospheric conditions present around sunspots. Our models use the Fokker-Planck approximation and employ a return current (\citealt{Holman:2012aa}). RADYN solves the non-LTE population densities for the first 6 levels of the hydrogen atom, the first 9 levels of the helium atom, and the first 6 levels of the calcium atom and computes the line profiles of  bound-bound and bound-free transitions within the atomic configuration described above. Complete frequency redistribution is considered for the line transitions, which may inhibit its ability to reproduce the wings of resonance lines.

We generated a grid of electron beam-driven models with $\delta$=3, F$=$3$\times$10$^{9}$, 10$^{10}$ and 3$\times$10$^{10}$~erg~cm$^{-2}$~s$^{-1}$ and $E_C$= 20, 40, 60, 80, 100 and 120~keV. Proton beams of the same flux and $\delta$ were modeled with $E_C$= 2, 4, 8, and 16~MeV. Models with F=10$^{9}$~erg~cm$^{-2}$~s$^{-1}$ did not show any detectable increase in the optical continuum and their output is not presented in this work. 

For the modeling of the higher order Balmer lines we used the non-LTE radiative transfer code RH (\citealt{Uitenbroek:2001aa}) with the latest modifications introduced by \cite{Kowalski:2017aa}. The code employs a 20 level hydrogen atom and models more accurately the electric pressure hydrogen line broadening that occurs in flares. RH also uses partial redistribution which is needed for accurate modeling of the \ion{Ca}{2}~K\&H line profiles. All simulations lasted 60~s with a rapid onset of the beam (0.1~s) applied until t~=~30~s, followed by a 3~s linear decay. The remaining 27~s had no beam heating applied, allowing the atmosphere to relax. 

\section{Results}
\label{results_section}
The observations described in Section \ref{flare_obs} allowed us to obtain a set of criteria that we applied on the grids of models. Using the IS data we obtain the constraint that the  H$\gamma$ vs. \ion{Ca}{2}~K line ratio should not be greater than 0.1. RHESSI and Fermi provided energetic criteria (Tables~\ref{tabX} and \ref{tabXF}) while the detection of positive WL contrast by SDO/HMI allow us to eliminate those models that do not show any positive WL contrast at 615 nm. 
\subsection{Electron Beam Models}
\label{ebeam_models}

\begin{figure}[h]
\centering
\includegraphics[width=0.75\textwidth]{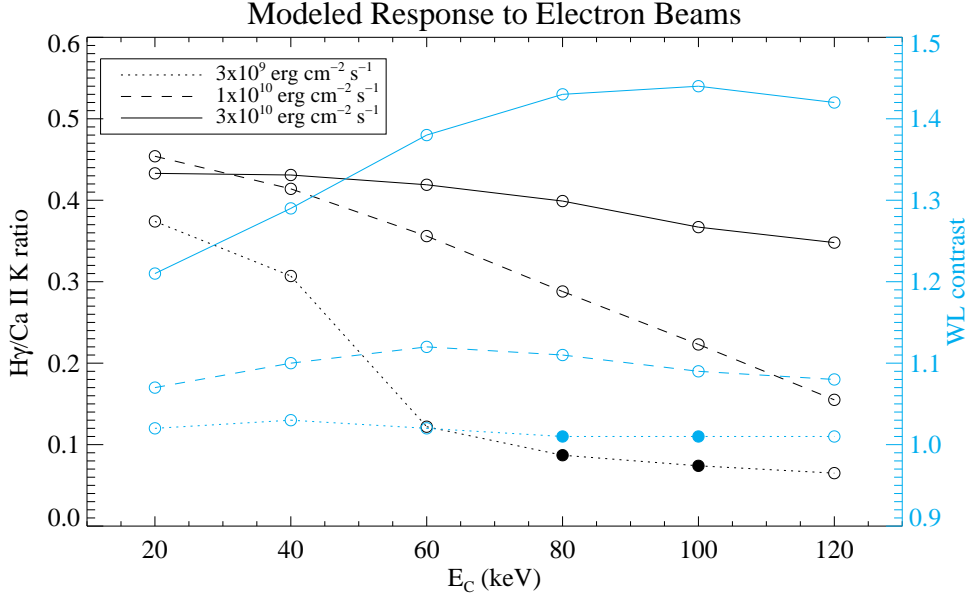}
\caption{H$\gamma$/\ion{Ca}{2}~K ratio (black) and WL contrast (cyan) as a function of $E_C$, and for different non-thermal electron fluxes (solid curves F=3E10, dashed curves F=1E10, dotted curves F=3E9). The solid circles denote the values imposed by the observational constraints.}
\label{etest_phase}
\end{figure}

We compared the synthetic spectra from RH with the observations. We focused our analysis on two continuum measurements, one redwards of the Balmer jump (364.7~nm) and another close to the HMI working wavelength (615~nm) as well as measurements of the H$\gamma$ core positions.  The continuum in the vicinity of the Balmer jump remained unchanged for a flux equal to 3$\times$10$^{9}$~erg~cm$^{-2}$~s$^{-1}$ and E$_C$ in a range of 60 to 100~keV. Higher beam fluxes resulted in an elevated continuum. The continuum at 615~nm showed the same trend with a higher contrast. Table~\ref{tab1} and Figure~\ref{etest_phase} show that the contrast of WL continuum peaks at greater $E_C$ values in models where the flux is greater. Broader line profiles were  found for models with higher energy flux and higher E$_C$ while the line strength decreased with increasing $E_C$. H$\gamma$ is included in the RADYN simulation output and has good counting statistics in the observational dataset, so it was chosen as a representative for the study of the higher order Balmer lines.
The rise in the H$\gamma$ was calculated as the ratio between flare and quiescent profiles in the wavelength range 434.159--434.186~nm. 
For all models we investigated the temperature in the upper photosphere (z=300~km) and the penetration depth of the beam (Table~\ref{tab1}). The penetration depth was defined as the range of heights above the photospheric floor where the volumetric beam heating reached at least 10\% of its maximum. The response of the H$\gamma$/\ion{Ca}{2}~K ratio and the WL contrast are plotted in Figure~\ref{etest_phase} as a function of $E_C$.

\begin{deluxetable}{cccccccccc}[h]
\tablecaption{Spectral diagnostics from electron beam-driven models 20 seconds into the simulation for a range of electron beam fluxes (F) and $E_C$. The signal presented in each waveband is a pure flaring signal relative to the initial/quiescent state. The models that comply to observations are typed in boldface. \label{tab1}}
\tablehead{& & $E_C$  & Temperature at  & Penetration & Rise in cont.  &  H$\gamma$/\ion{Ca}{2} K  & Rise in & Rise in cont. \\
         & &  (keV) & z = 300 km (K)  &  depth (km) & $>$ 364.7 nm &  ratio & core of H$\gamma$ & 615 nm}
\startdata
\multirow{6}{*}{\rotatebox{90}{F = 3$\times$10$^{9}$}} & \multirow{6}{*}{\rotatebox{90}{erg cm$^{-2}$ s$^{-1}$}} & 20 &  4608 &  480--1158 &  1.01 &  0.374 &  9.18 &  1.02 \\
&& 40 &  4677 &  410--1156 &  1.01 & 0.307  &  6.53 &  1.03 \\
&& 60 &  4630 &  363--1118 &  1.00 & 0.122 &  3.45 &  1.02 \\
&& {\bf 80} &  {\bf 4617} & {\bf  340--1040}& {\bf  1.00} & {\bf 0.087}&{\bf   2.63} & {\bf  1.01} \\
&& {\bf 100} & {\bf  4614} & {\bf  316--997}& {\bf  1.00 }&{\bf 0.074} &{\bf   2.25} &{\bf   1.01} \\
&& 120 &  4614 &  293--975 &  0.99 & 0.065  &  2.02 &  1.01 \\
\hline
\multirow{6}{*}{\rotatebox{90}{F = 1$\times$10$^{10}$ }} & \multirow{6}{*}{\rotatebox{90}{erg cm$^{-2}$ s$^{-1}$}} & 20 &  4723. &  433--1164 &  1.03 &  0.454 & 16.79 &  1.07 \\
&&40 &  4811 &  433--1156 &  1.04 & 0.414  & 14.36 &  1.10 \\
&&60 &  4990 &  363--1156 &  1.05 & 0.356  & 10.11 &  1.12 \\
&&80 &  4923 &  316--1149 &  1.04 & 0.288&  7.68 &  1.11 \\
&&100 &  4873 &  293--1040 &  1.03 & 0.223  &  6.19 &  1.09 \\
&&120 &  4844 &  293--975 &  1.03 & 0.155  &  4.58 &  1.08 \\
\hline
\multirow{6}{*}{\rotatebox{90}{F = 3$\times$10$^{10}$}} & \multirow{6}{*}{\rotatebox{90}{erg cm$^{-2}$ s$^{-1}$}} & 20 &  5022. &  640--1167 &  1.11 & 0.433  & 25.54 &  1.21 \\
&&40 &  5336 &  387--1156 &  1.15 & 0.431  & 24.74 &  1.29 \\
&&60 &  5393 &  410--1156 &  1.19 & 0.419 & 22.15 &  1.38 \\
&&80 &  5489 &  340--1156 &  1.22 & 0.399  & 17.73 &  1.43 \\
&&100 &  5642 &  316--1156 &  1.23 & 0.367  & 14.39 &  1.44 \\
&&120 &  5571 &  293--1147 &  1.21 & 0.348  & 12.22 &  1.42 \\
\enddata
\end{deluxetable}

\begin{figure}[h]
\centering
\includegraphics[width=0.75\textwidth]{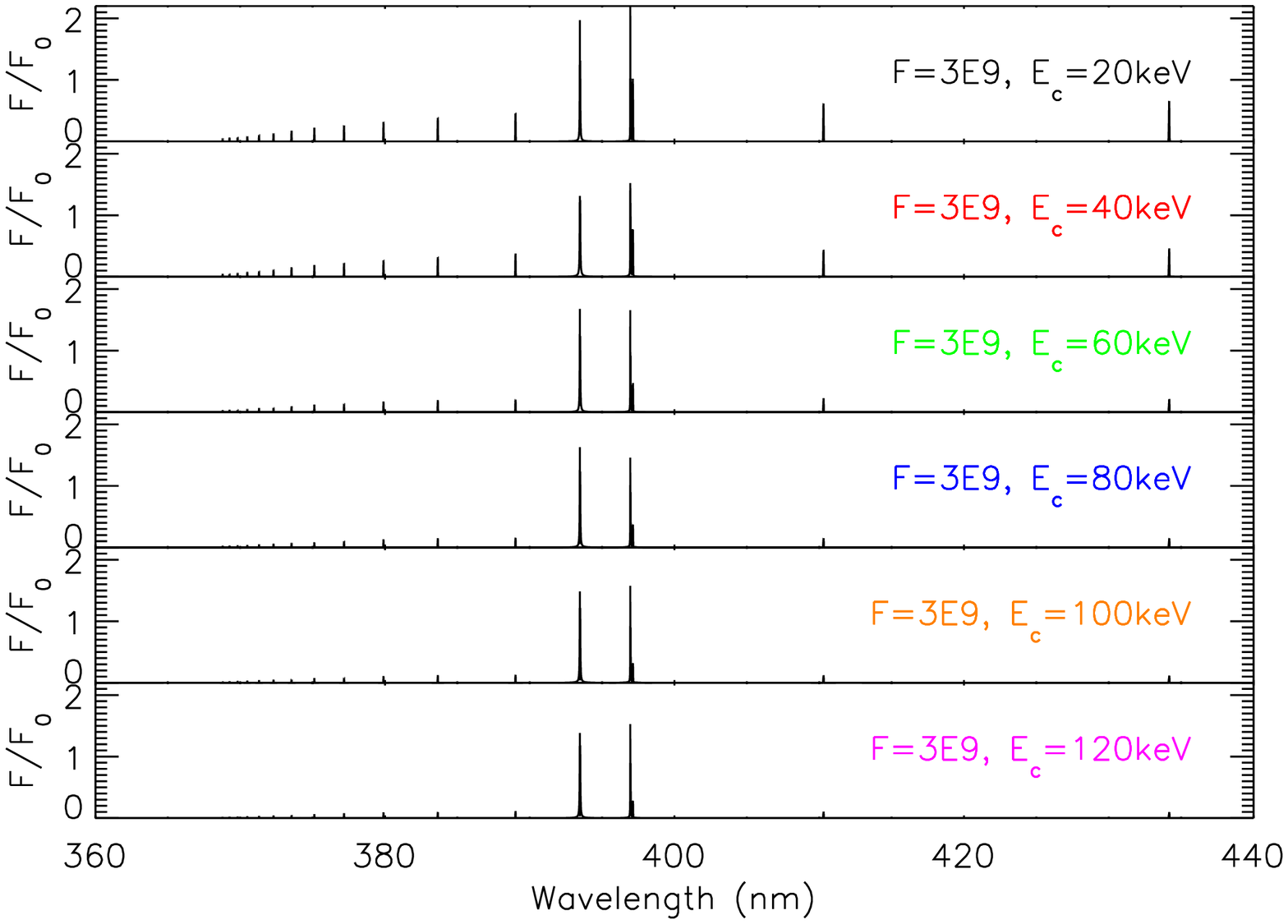}
\caption{Relative flare excess for electron beam-driven models for a filling factor of 0.07.  All higher order Balmer lines become weaker with increasing E$_C$. The \ion{Ca}{2} K\&H lines at 393.4 and 396.8 nm are also shown.}
\label{fig6}
\end{figure}

\begin{figure}[h]
\centering
\includegraphics[width=0.85\textwidth]{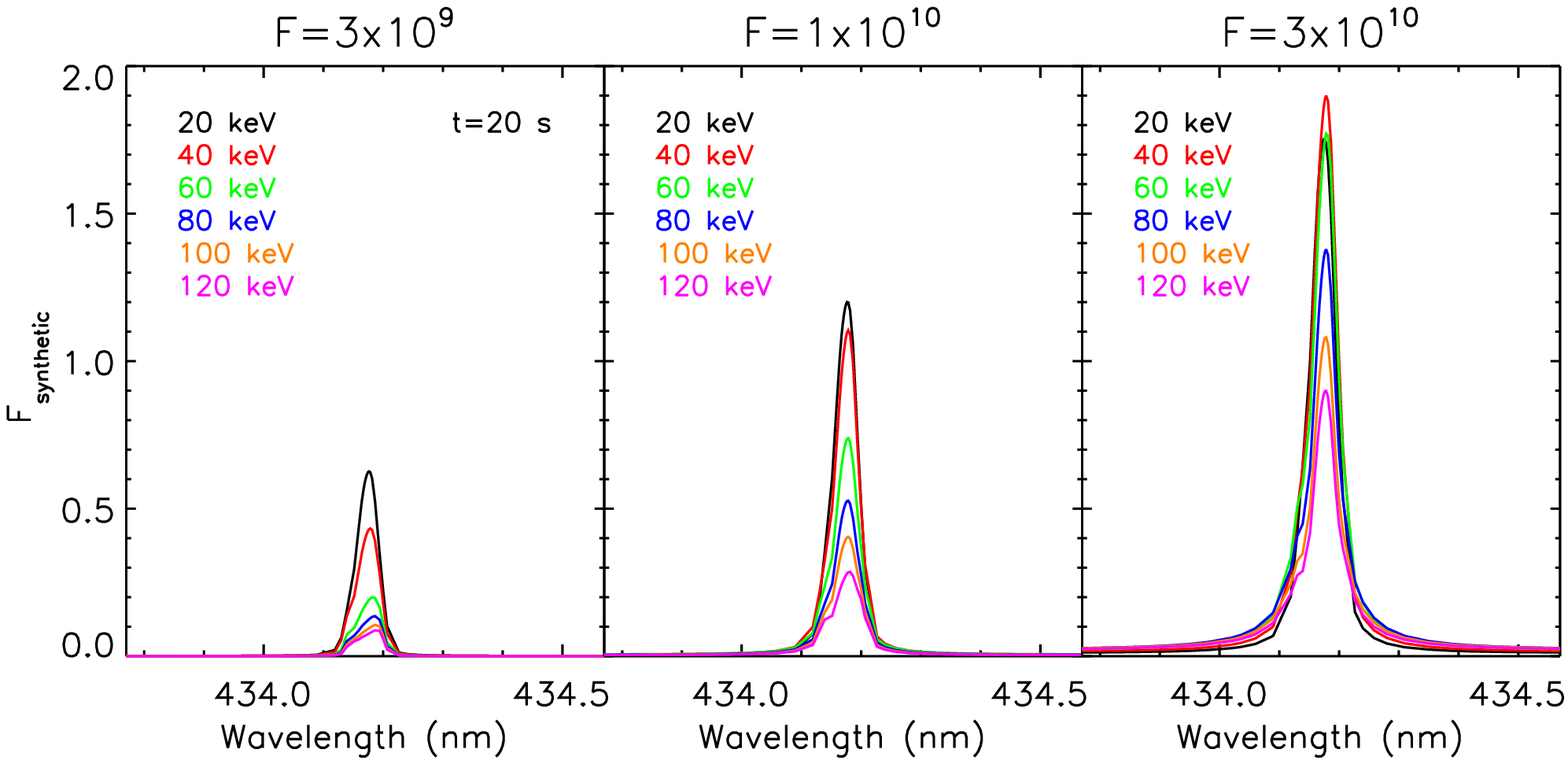}
\caption{A detail of H$\gamma$ line for electron beam fluxes of 3$\times$10$^{9}$, 10$^{10}$ and 3$\times$10$^{10}$~erg~cm$^{-2}$~s$^{-1}$ with filling factor of 0.07. Legend shows color coded values of the $E_C$. The flaring spectra were takes at 20 s into the simulations.}
\label{fig2}
\end{figure}

In order to compare our models with observations in the visible/NUV, we combined the flare signal ($F_{flare}$) obtained at t=20~s into the simulation with the non-flare signal ($F_{non-flare}$) obtained at t=0~s. Then we subtracted the non-flare signal to obtain the flare excess and divided it with the non-flare signal 
\begin{equation}
F_{synthetic}=\frac{F_{flare}*ff+F_{non-flare}*(1-ff)-F_{non-flare}}{F_{non-flare}}.
\end{equation}
The electron beam-driven models with a flux of 3$\times$10$^{9}$~erg~cm$^{-2}$~s$^{-1}$ and $E_C$ between 60 and 100~keV showed a positive WL contrast at 615~nm and lie within the observed range of energies from RHESSI. These beams produced H$\gamma$ to \ion{Ca}{2} K ratios of 0.122 (60~keV), 0.087 (80~keV) and 0.074 (100~keV) assuming $f\!f$= 0.07 (Figure~\ref{fig6}).
We found that any emission in the higher order Balmer lines was below the noise levels and could not be detected in the observed spectra. The 60, 80 and 100~keV (F = 3$\times$10$^{9}$~erg~cm$^{-2}$~s$^{-1}$) electron beams produce very weak WL contrast of 1.02, 1.01 and 1.01, respectively. Models with the same flux but lower $E_C$, also produce too strong emission in the H$\gamma$ line (Figure \ref{fig2}). A higher energy flux (1$\times$10$^{10}$~erg~cm$^{-2}$~s$^{-1}$ or more) is consistent with the RHESSI constraints only for E$_C$=20 keV (Table~\ref{tabX}), but such a beam triggers too strong emission in H$\gamma$ (Figure~\ref{fig2} and Table~\ref{tabX}).

\subsection{Proton Beam Models}
\label{pbeam_models}
\begin{figure}[h]
\centering
\includegraphics[width=0.75\textwidth]{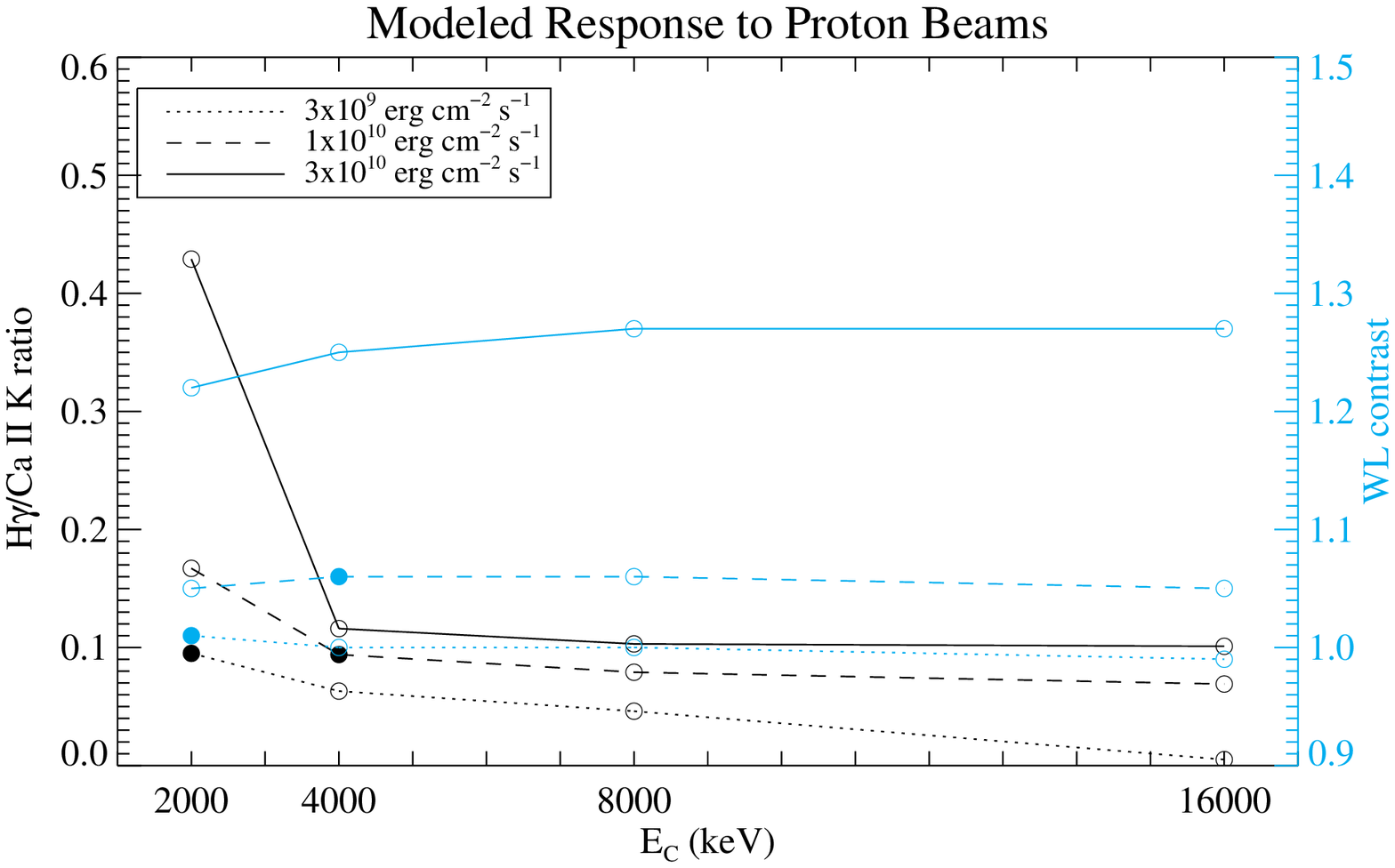}
\caption{H$\gamma$/\ion{Ca}{2}~K ratio (black) and WL contrast (cyan) as a function of $E_C$, and for different non-thermal proton fluxes (solid curves F=3E10, dashed curves F=1E10, dotted curves F=3E9). The solid circles denote the values allowed by the data constraints.}
\label{ptest_phase}
\end{figure}

\begin{figure}[h]
\centering
\includegraphics[width=0.85\textwidth]{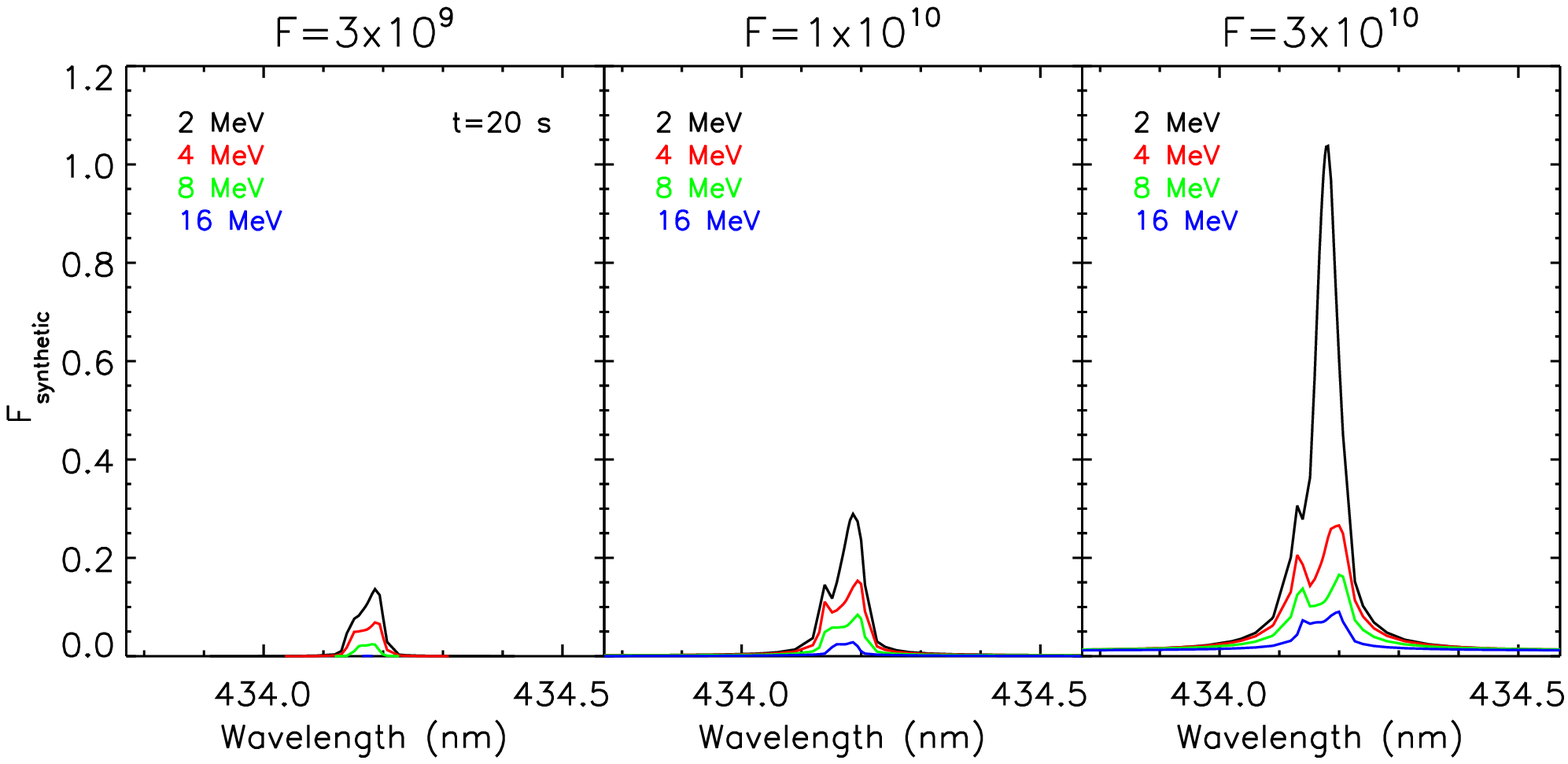}
\caption{A detail of H$\gamma$ line for proton beam fluxes of 3$\times$10$^{9}$, 10$^{10}$ and 3$\times$10$^{10}$~erg~cm$^{-2}$~s$^{-1}$ with filling factor of 0.07. Legend shows color coded values of the E$_C$. The flaring spectra were takes at 20 s into the simulations.}
\label{fig5}
\end{figure}

The outputs from the RADYN/RH models of proton beam heating for a range of low-energy cutoff and proton fluxes are given in Figure~\ref{ptest_phase} and Table~\ref{tab2}.
The H$\gamma$ line of the proton beam-driven models in Figure~\ref{fig5} shows a similar pattern to electron beam-driven models with more pronounced central reversal, however in general the line tends to be weaker for proton beams.

\begin{figure}[h]
\centering
\includegraphics[width=0.75\textwidth]{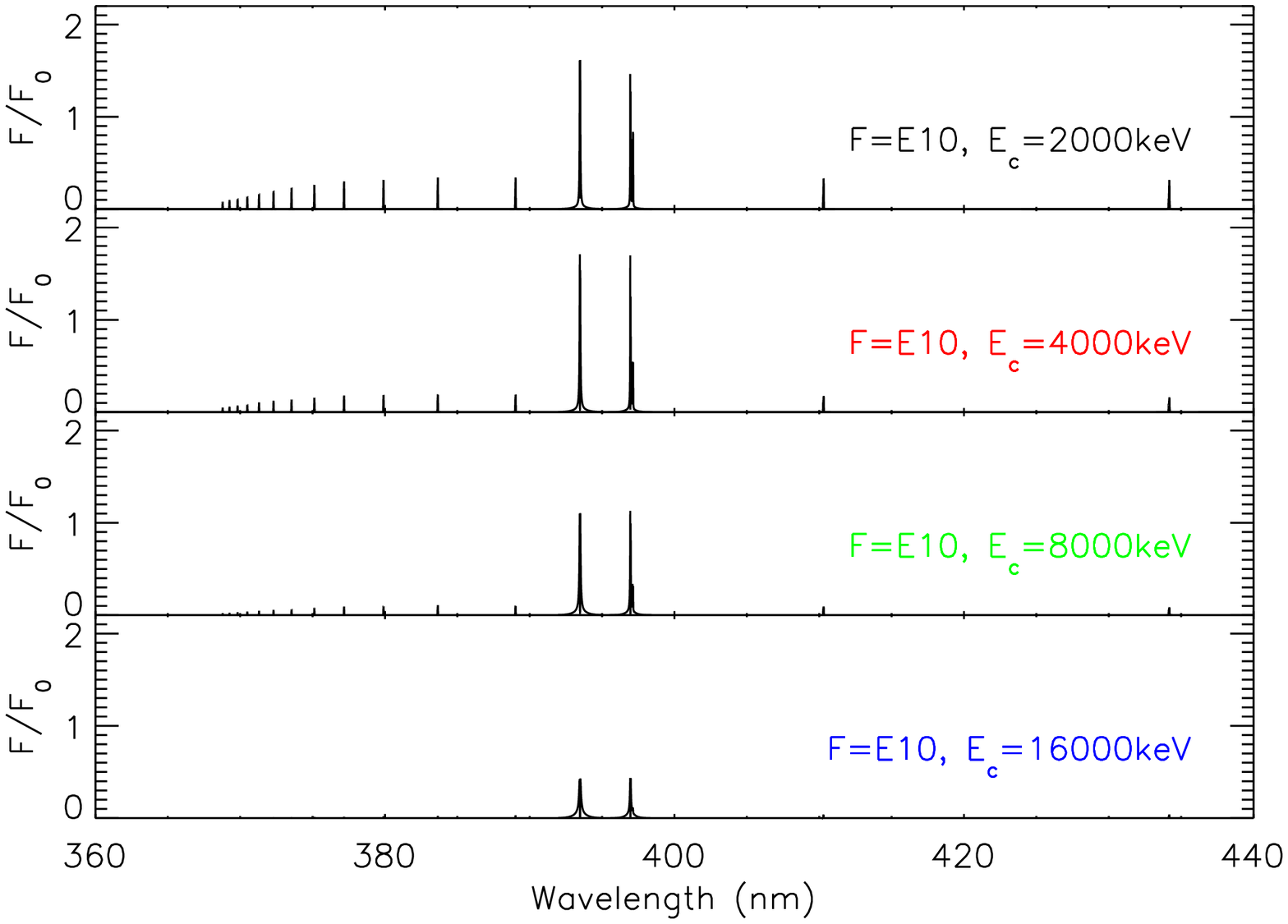}
\caption{Relative flare excess for proton beam-driven models. Filling factor = 0.07.}
\label{fig7}
\end{figure}

Of the modeled proton beams, a flux equal to 3$\times$10$^{9}$~erg~cm$^{-2}$~s$^{-1}$ produced a positive WL contrast only for $E_C$=2~MeV,  where the H$\gamma$ vs. \ion{Ca}{2}~K line ratio was equal to 0.095. For a beam with a flux of 1$\times$10$^{10}$~erg~cm$^{-2}$~s$^{-1}$ Fermi detected sufficient power if $E_C\le$3.8~MeV. We do not expect any significant differences of this beam from the $E_C$=4~MeV beam that we modeled. Table \ref{tab2} and Figure \ref{fig7} show that for this model, the H$\gamma$ vs. \ion{Ca}{2}~K line ratio reached 0.094, which is very close to the 3$\times$10$^{9}$~erg~cm$^{-2}$~s$^{-1}$ proton beam-driven model mentioned above. A flux equal to 3$\times$10$^{10}$~erg~cm$^{-2}$~s$^{-1}$ produced too strong emission in the higher order Balmer lines ($E_C$~=~2~MeV) or its energy was out of the range observed by Fermi ($E_C\ge$~4~MeV).

\begin{deluxetable}{cccccccccc}[h]
\tablecaption{Spectral diagnostics in proton beam-driven models 20 seconds into the simulation for a range of proton beam fluxes and $E_C$. The signal presented in each waveband is  a pure flaring signal relative to the initial/quiescent state. The models that comply with the observations are typed in bold.  \label{tab2}}
\tablehead{&&$E_C$  & Temperature at   & Penetration & Rise in cont.  & H$\gamma$/\ion{Ca}{2} K  & Rise in  & Rise in cont. \\
         && (keV) &  z = 300~km (K) &  depth (km) & $>$ 364.7~nm &  ratio & core of H$\gamma$ &  615~nm}
\startdata
\multirow{4}{*}{\rotatebox{90}{F = 3$\times$10$^{9}$}} & \multirow{4}{*}{\rotatebox{90}{erg cm$^{-2}$ s$^{-1}$}} &    {\bf 2000} &  {\bf 4598} &  {\bf 293--1078} &  {\bf 0.99} &  {\bf 0.095} &  {\bf 2.45} &  {\bf 1.01} \\
 &&   4000 &  4621 &  172--907 &  0.99 & 0.063 &  1.77 &  1.00 \\
 &&   8000 &  4659 &   73--753 &  0.98 & 0.046  &  1.31 &  1.00 \\
 &&  16000 &  4713 &  -46--618 &  0.97 & 0.005  &  1.01 &  0.99 \\
\hline
\multirow{4}{*}{\rotatebox{90}{F = 1$\times$10$^{10}$ }} & \multirow{4}{*}{\rotatebox{90}{erg cm$^{-2}$ s$^{-1}$}} &    2000 &  4730. &  293--1093 &  1.01 &  0.167 &  3.85 &  1.05 \\
 &&   {\bf 4000} & {\bf  4803} &  {\bf 172--907} &  {\bf 1.00} & {\bf 0.094} & {\bf  2.53} &  {\bf 1.06} \\
 &&   8000 &  4907 &   49--753 &  0.99 &  0.079 &  1.86 &  1.06 \\
 &&  16000 &  5018 &  -46--618 &  0.97 & 0.069  &  1.35 &  1.05 \\
\hline
\multirow{4}{*}{\rotatebox{90}{F = 3$\times$10$^{10}$}} & \multirow{4}{*}{\rotatebox{90}{erg cm$^{-2}$ s$^{-1}$}} &    2000 &  5195. &  221--1149 &  1.09 & 0.429  & 11.59 &  1.22 \\
 &&   4000 &  5292 &  172--929 &  1.10 & 0.116  &  3.70 &  1.25 \\
 &&   8000 &  5440 &   49--753 &  1.10 & 0.103 &  2.54 &  1.27 \\
 &&  16000 &  5591 &  -53--618 &  1.08 &  0.101 &  1.99 &  1.27 \\
\enddata
\end{deluxetable}

\section{Discussion}
\label{disc_section}

The X1 flare event presented in this work showed an extremely hard electron spectrum ($\delta \approx 3$), which makes it difficult to estimate an accurate value for $E_C$. In such a hard spectrum the flattening of the photon spectrum below $E_C$ is likely to be the same regardless of the value of $E_C$ (see Section \ref{rhessi_section}). Our analysis has produced two electron and two proton beam-driven models that can reproduce the observations within our observational  uncertainties. Of these four models the 1$\times$10$^{10}$~erg~cm$^{-2}$~s$^{-1}$ and 4~MeV proton beam produced the highest WL contrast (1.06) that agrees best with the observed value. The other three models had a lower beam flux and showed a WL contrast of only 1.01. There was no increase detected redward of the Balmer jump ruling out a presence of a hot black-body component in the photosphere and agrees with the observations. Further examination showed that electrons do not penetrate as deep as protons (316 vs. 172~km above the photospheric floor), which is consistent with a lower temperature in upper photosphere (4614 vs. 4803~K). The models also show that electron beams deliver their energy in the upper chromosphere ($\sim$1150 km) for the beams within the energy range allowed by RHESSI, except of that with F~=~3$\times$10$^{9}$~erg~cm$^{-2}$~s$^{-1}$ and $E_C\geq$ 80~keV. For proton beams it is easier to penetrate through the upper chromosphere without depositing a significant amount of energy there.

We see similar observational trends for both the proton and electron beam-driven models - the WL emission gets stronger with increasing beam flux, but this also triggers stronger emission in the Balmer lines. With increasing $E_C$ the emission in the Balmer lines becomes significantly weaker, while the WL emission at 615~nm shows relatively small changes (see Tables \ref{tab1} and \ref{tab2}).

\begin{figure}[h]
\centering
\includegraphics[width=0.45\textwidth]{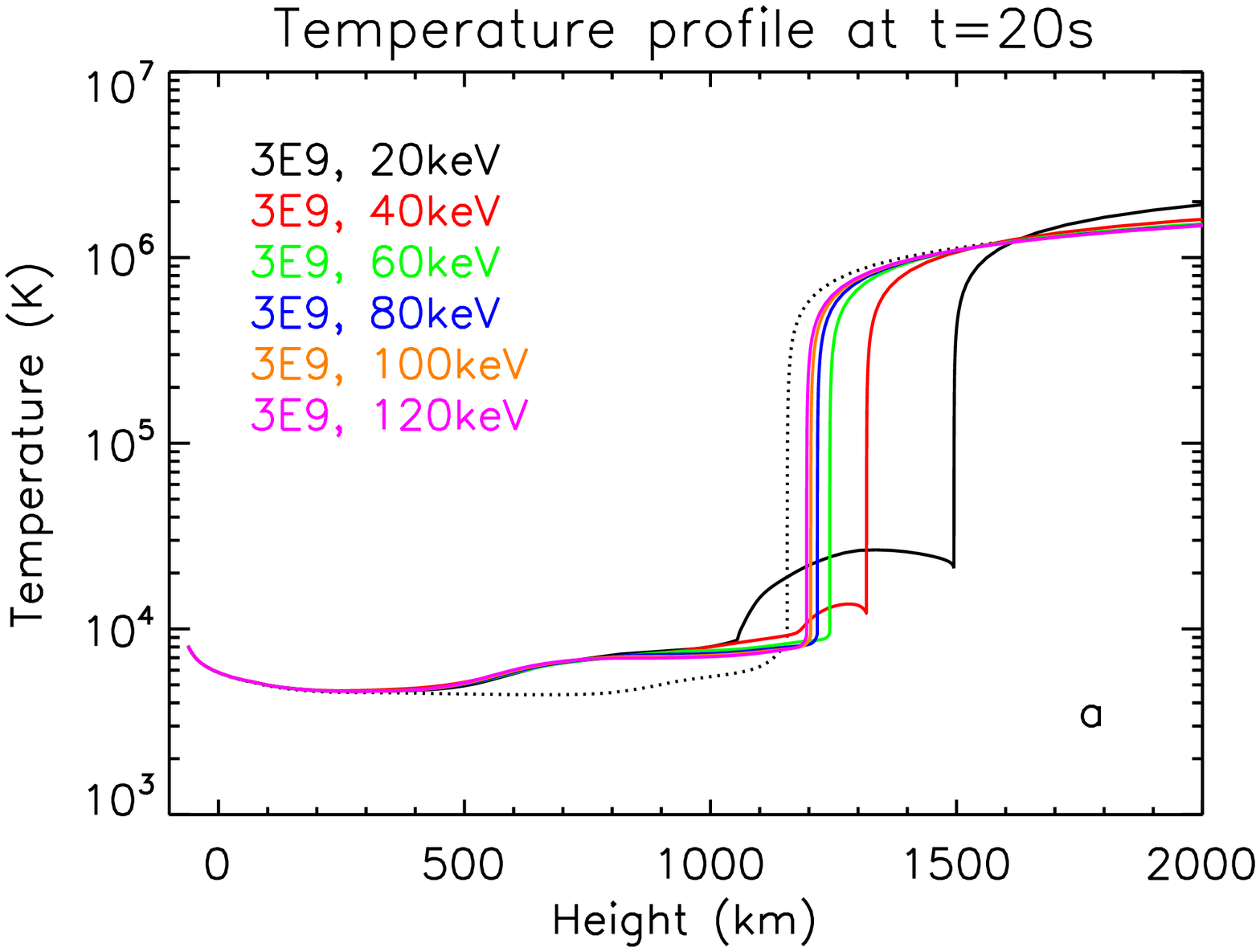} 
\includegraphics[width=0.45\textwidth]{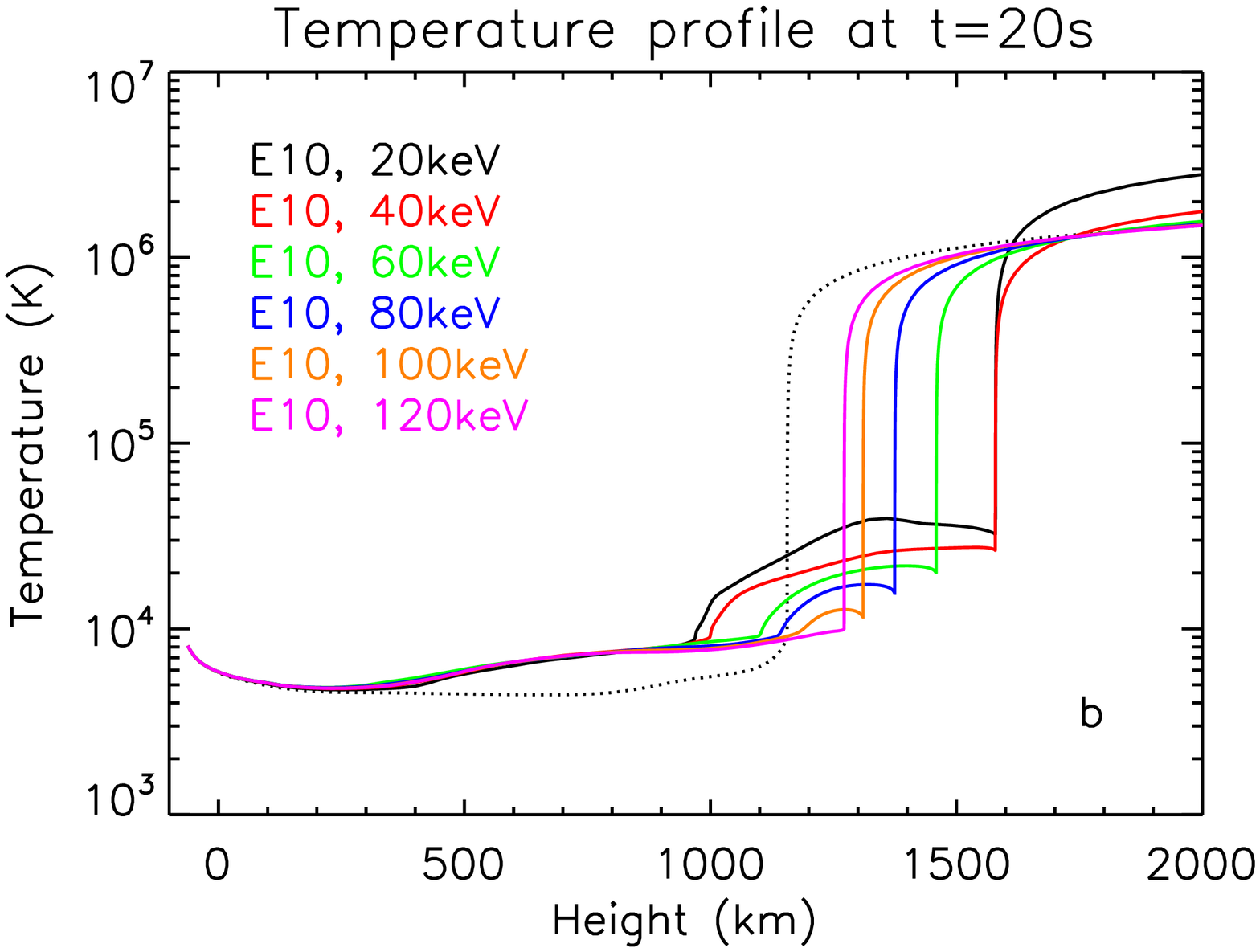}
\includegraphics[width=0.45\textwidth]{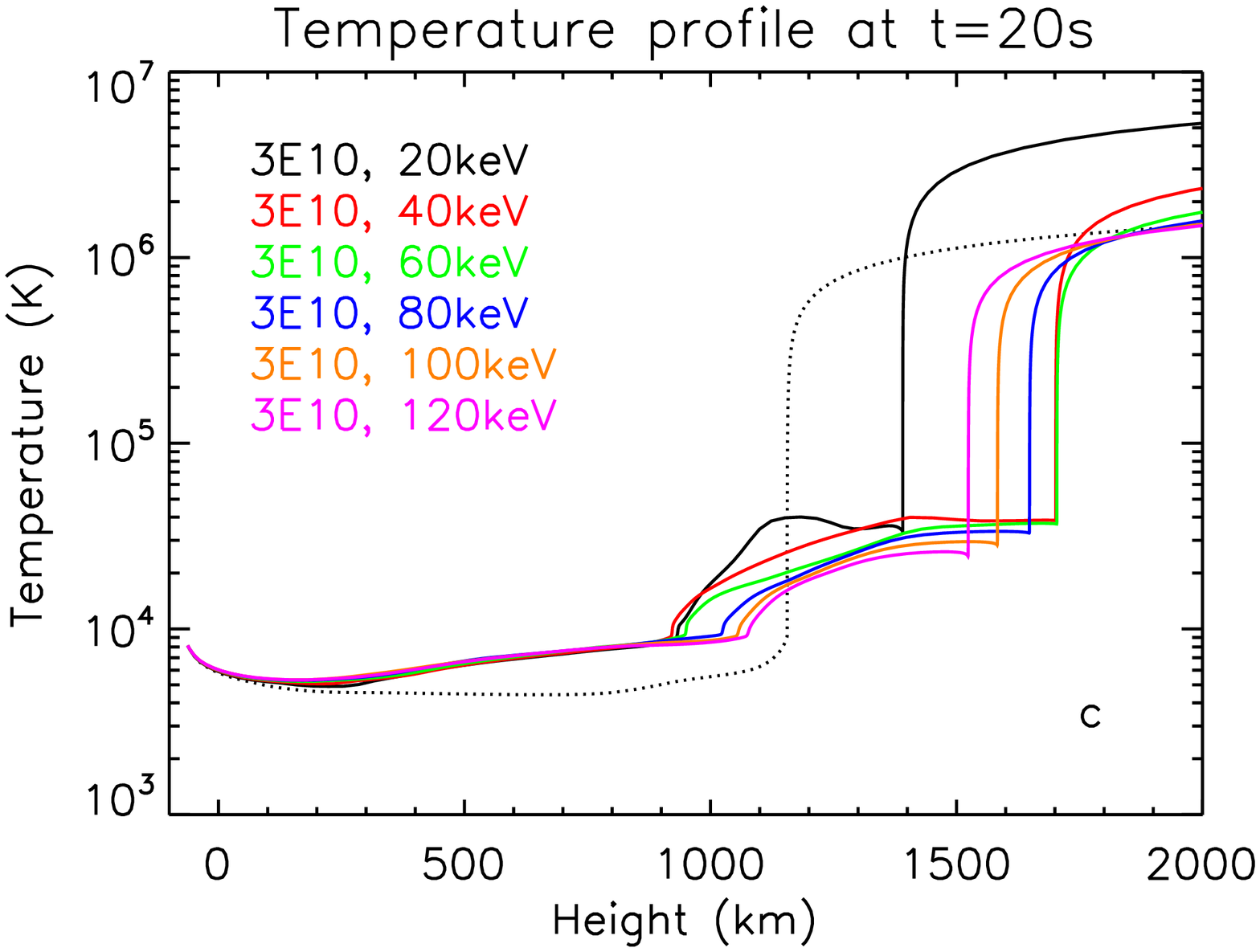}
\includegraphics[width=0.45\textwidth]{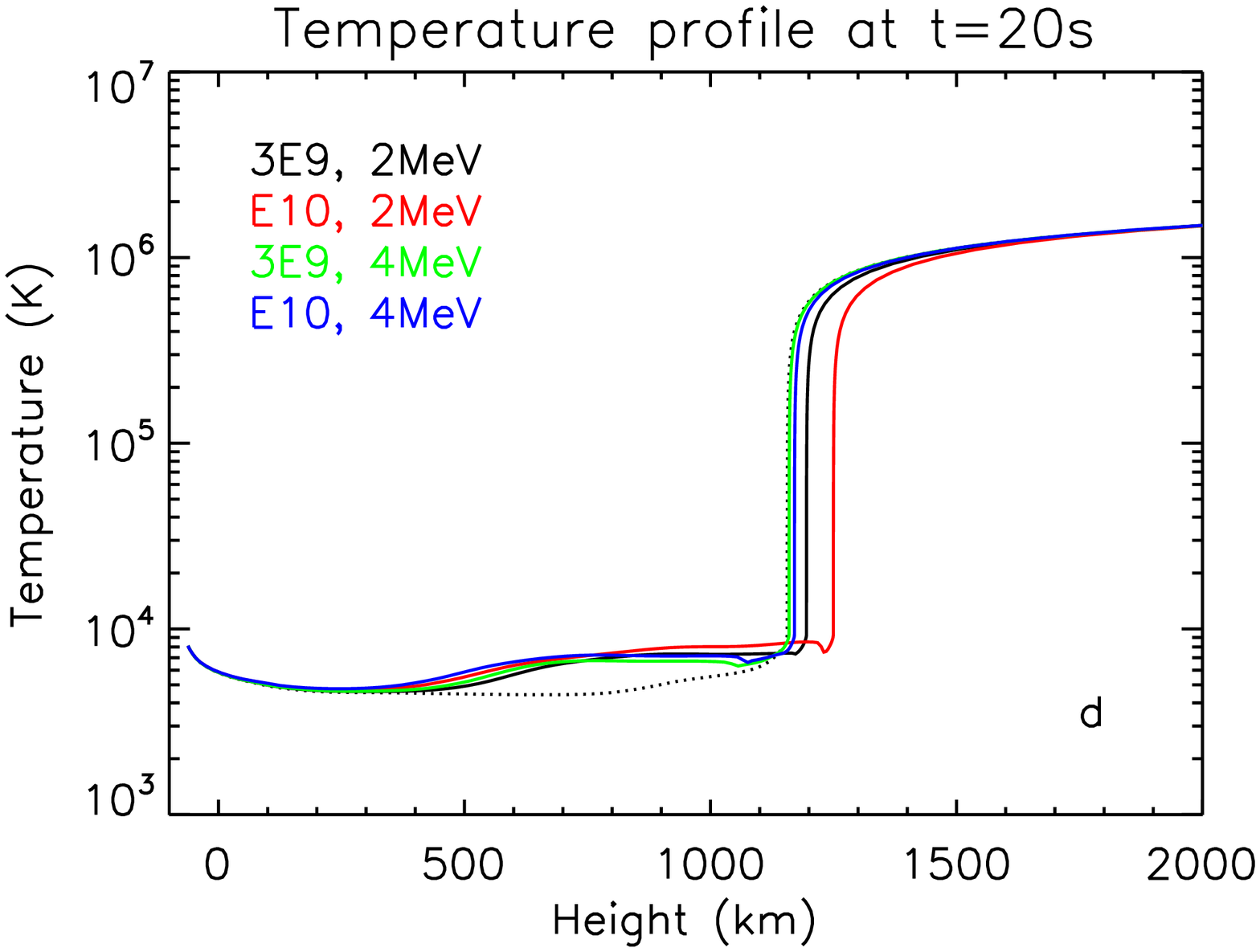} 
\caption{Temperature profiles for the electron beam-driven models (panels a, b, and c) and proton beam-driven models (panel d). The dotted line indicates the initial pre-flare atmosphere.}
\label{fig4}
\end{figure}

The temperature profiles in Figure~\ref{fig4} show that the electron beams with a flux of at least 1$\times$10$^{10}$~erg~cm$^{-2}$~s$^{-1}$ cause a large disturbance in the chromosphere and shift the transition region to greater geometrical heights. This is always accompanied with emission in the Balmer lines. In contrast, the proton beam-driven models with the same flux leave the upper chromosphere relatively undisturbed. For both electron and proton beam-driven models a temperature rise appears deeper for higher fluxes, while the value of E$_C$ has only a minor effect. This supports the idea that WLFs have a lower limit to their total flux, because only beams which are sufficiently intense can 
trigger emission in the continuum (615~nm, 7th column of Tables~\ref{tab1} and \ref{tab2}). This also contradicts the statement by \cite{Jess+all2008} that `there is no reason why WLF emission should not be produced in all flares'. However, the value of $E_C$ plays a significant role on the effects of the beam in the chromosphere. For low values of the $E_C$, the beam delivers its energy in the upper chromosphere where the Balmer lines are formed. We found that only those electron beams with E$_C$ exceeding the range of modeled values (20--120~keV) dissipate the energy deep in the atmosphere and keep the Balmer lines in absorption. Our result does not agree with the conclusion of \cite{Fletcher:2007ab} that the visible/UV continuum requires an electron beam with a cut-off energy well below 25 keV in order to deliver sufficient energy into the atmosphere, nor with the commonly observed values of $E_C$ reported by \cite{Fletcher:2007aa}. The power in non-thermal electrons was 1.59$\times$10$^{27}$~erg~s$^{-1}$ when considering $E_C$=20 keV, which is at least an order of magnitude lower than in the average WLF (\citealt{Watanabe:2017aa}). As far as we know there in no work presenting quantitative analysis on proton beams in WLFs, but the deposition rate gives preference to proton beams (P$_{E_p>2MeV}$=7.1$\times$10$^{27}$~erg~s$^{-1}$) to be the main driver of the WL emission in this flare. 

\begin{figure}[h]
\centering
\includegraphics[width=0.5\textwidth]{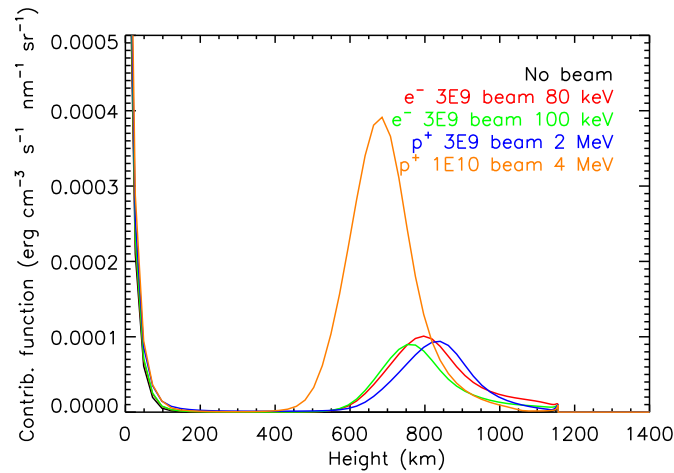}
\caption{Contribution functions to continuum at 615~nm of the favored electron (e$^-$) and proton (p$^+$) beam-driven models.}
\label{fig8}
\end{figure}

The contribution functions of the favored electron and proton beam-driven models are shown in Figure \ref{fig8} with the main parameters summarized in Table~\ref{tab_results}. It is defined by \cite{Carlsson:1997aa} as
\begin{equation}
I_{\nu}=\int_{z_0}^{z_1}S_{\nu}e^{-\tau_{\nu}}\chi_{\nu}dz,
\end{equation}
where $I_{\nu}$ is intensity at frequency $\nu$, $S_{\nu}$ is a source function, $e^{-\tau_{\nu}}$ is an exponential attenuation factor  and $\chi_{\nu}$ is the monochromatic opacity.
The beam flux appeared to be the most important factor with respect to both origin of the WL excess emission and WL contrast. A comparison of electron beams with $E_C$ of 80 and 100~keV indicates that higher values of $E_C$ result in a deeper penetration of the beam however, flux plays a more important role. From the numerical results it is clear that the photospheric contribution (defined as the contribution function integrated over heights 0 - 300 km) to the WL excess emission plays a minor role in this event and the excess continuum at 615~nm is predominantly formed in the lower chromosphere, no matter whether it is driven by an electron or proton beam. The simulations do not show any significant shifting of the $\tau$=1 surface for the modeled beams. We therefore conclude that the excess WL emission is optically thin resulting in the minor WL enhancement detected in this event.

\begin{deluxetable}{cccccc}[h]
\tablecaption{A summary of the beam parameters that are in best agreement with the observations. For WL emission only the excess is quantified (4th, 5th and 6th columns). \label{tab_results}}
\tablehead{Flux  & E$_c$  &   H$\gamma$/Ca~II~K & WL  & Origin of middle 90\% WL  &  Photospheric\\
          (erg cm$^{-2}$ s$^{-1}$) & (keV) &   ratio & contrast & (height above photosphere) & contribution}
\startdata
    (e$^-$) 3$\times$10$^{9}$ & 80 &  9 \% & 1 \% &  663--1060 km & 4.1 \%\\
    (e$^-$) 3$\times$10$^{9}$ & 100 &  8 \% & 1 \% &  640--1040 km & 7.0 \%\\
    (p$^+$) 3$\times$10$^{9}$ & 2000 &   10 \% & 1 \% &  663--1019 km & 14.2 \%\\
    (p$^+$) 1$\times$10$^{10}$ & 4000 &   9 \% & 6 \% &  559--863 km & 6.2 \%\\
\enddata
\end{deluxetable}

\section{Concluding remarks}
\label{conc_section}

The main motivation behind this work has been to understand the nature of the suppressed Balmer line emission observed in the 2014 June 11 X1.0 white light flare. We used particle beam parameters constrained from RHESSI and Fermi spectra to generate a number of RHD models. Our models have shown that the spectral signatures of Type II WLF can be best reproduced with a relatively weak particle beam that has a high low energy cut-off (Table \ref{tab_results}). Beams with such parameters in X-class solar flares are rare (\citealt{Kuhar:2016aa}). Our models also show that both electron and proton beams can be responsible for Type II WLF, but proton beams penetrate more easily through the upper chromosphere without triggering a strong emission in the higher order Balmer lines and at the same time can carry more energy. The excess WL emission then originates over a broad range of heights in the lower chromosphere with a relatively small contribution from the photosphere. We found that solely based on a match between the WL emission and the peak of HXR, we cannot decide if the studied event is a Type I or Type II WLF, as \cite{Metcalf:2003aa} did.

One of the limitations of this work is that, due to the nature of the RADYN 1D geometry, our modeling approach is more accurate when the line-of-sight does not deviate significantly from the loop axis. An off-axis line-of-sight requires a 3D RHD model to account for the overlying non-flaring chromosphere. For the evaluation of the spectra we use the relative heights of flare excess emission in H$\gamma$ and \ion{Ca}{2} K lines. As the cores of both lines are formed at similar atmospheric heights, we assume that the overlying atmosphere would have similar effects to cores of both lines.\\
Notwithstanding that Alfv\'{e}n waves cannot be ruled out as drivers of the WL emission, the present paper only focuses on electron and proton beams as the version of RADYN used in our work does not allow us to investigate this scenario.

\acknowledgments
The research leading to these results has received funding from the European Community's Seventh Framework Programme (FP7/2007-2013) under grant agreement no. 606862 (F-CHROMA). R.O.M. would like to acknowledge support from NASA LWS/SDO Data Analysis grant NNX14AE07G, and the Science and Technologies Facilities Council for the award of an Ernest Rutherford Fellowship (ST/N004981/1). P.J.A.S. acknowledges support from the University of Glasgow's Lord Kelvin Adam Smith Leadership Fellowship. J.C.A acknowledges support from the Heliophysics Guest Investigator and Supporting Research Programs. 


\bibliography{big_lib}




\end{document}